\documentclass[
    ,final            
  ]
  {aipproc}

\layoutstyle{6x9}

\begin{document}

\title{Pion Interferemetry from p+p to Au+Au in STAR}

\classification{\texttt{25.75Gz}}

\author{Z. Chaj\c{e}cki (for the STAR  Collaboration)}
{  address={The Ohio State University, 191 W. Woodruff Avenue, Columbus, Ohio 43210, USA} }

\begin{abstract}
The geometric substructure of the particle-emitting source has been
characterized via two-particle interferometry by the STAR collaboration
for all energies and colliding systems at RHIC. We present systematic
studies of charged pion interferometry. The collective
nature of the source is revealed through the $m_T$ dependence of HBT
radii for all particle types. Preliminary results suggest a scaling in
the pion HBT radii with overall system size, as central Au+Au collisions
are compared to peripheral collisions as well as with Cu+Cu and even with d+Au and p+p
collisions, naively suggesting comparable flow strength in all systems.
To probe this issue in greater detail, multidimensional correlation
functions are studied using a spherical decomposition method.  This
allows clear identification of source anisotropy and, for the light
systems, the presence of significant long-range non-femtoscopic
correlations.
\end{abstract}

\maketitle


\section{Introduction}

Particle interferometry is an useful technique that provides information on 
the space-time properties of the particle emitting source and may be helpful 
in understanding the dynamics of the system created in high energy collisions 
by studying the transverse mass dependence ($m_{T} = \sqrt{ k_{T}^{2} + m_{\pi}^{2}}, k_{T}=\frac{1}{2}(p_{T_1}+p_{T_2})$ ) 
of HBT radii (for the latest review article see \cite{Lisa2005}). 
In this article femtoscopic results from a small system (p+p collisions) 
and a large system (Au+Au collisions) measured by the same experiment, at 
the same collision energy and  detector acceptance are presented 
for the first time in  high energy  physics.
The particular focus is on  the $m_T$ dependence of HBT radii  and an  attempt 
to understand its origins for different initial sizes of the emitting source is made.

\section{Analysis details}

The STAR Time Projection Chamber (TPC) \cite{TPC} was used to reconstruct particles of 
interest. Particle identification was achieved by measuring momentum and specific ionization losses of
charged particles in the gas of TPC ($dE/dx$ technique).
A large data statistics in Au+Au, Cu+Cu and d+Au collisions allows to do an analysis 
for different centralities. 
Additionally, d+Au data set allows to extract p+Au collisions. 
It is performed by selecting events with a single neutron tagged in 
the Zero Degree Calorimeter (ZDC) in the deuteron beam direction.

In this study pions with transverse momentum $0.10$ GeV/c $ <p_T< 1.00$ GeV/c were used and the 
analysis was done for four bins of $k_T$ within a range of $[0.15,0.60]$ GeV/c.
Two-track effects due to splitting (one particle reconstructed as two tracks)
 and merging (two particles reconstructed as one track) were removed from the data.

The dependence of the correlation function on the transverse momentum is studied 
as a function of three components of pair relative momentum in a Pratt-Bertsch coordinate 
system \cite{Pratt90,Bertsch89} in the longitudinally co-moving frame.
The fit was performed using a method suggested by Bowler \cite{Bowler} and Sinyukov \cite{Sinyukov}
assuming Gaussian parametrization of the source. For more details on analysis technique see \cite{Mercedes2004}.

\section{System Expansion and Multiplicity Scaling}
\begin{figure}[b]
\begin{minipage}[t]{73mm}
   \vspace*{-0.3cm}
  \includegraphics[width=73mm]{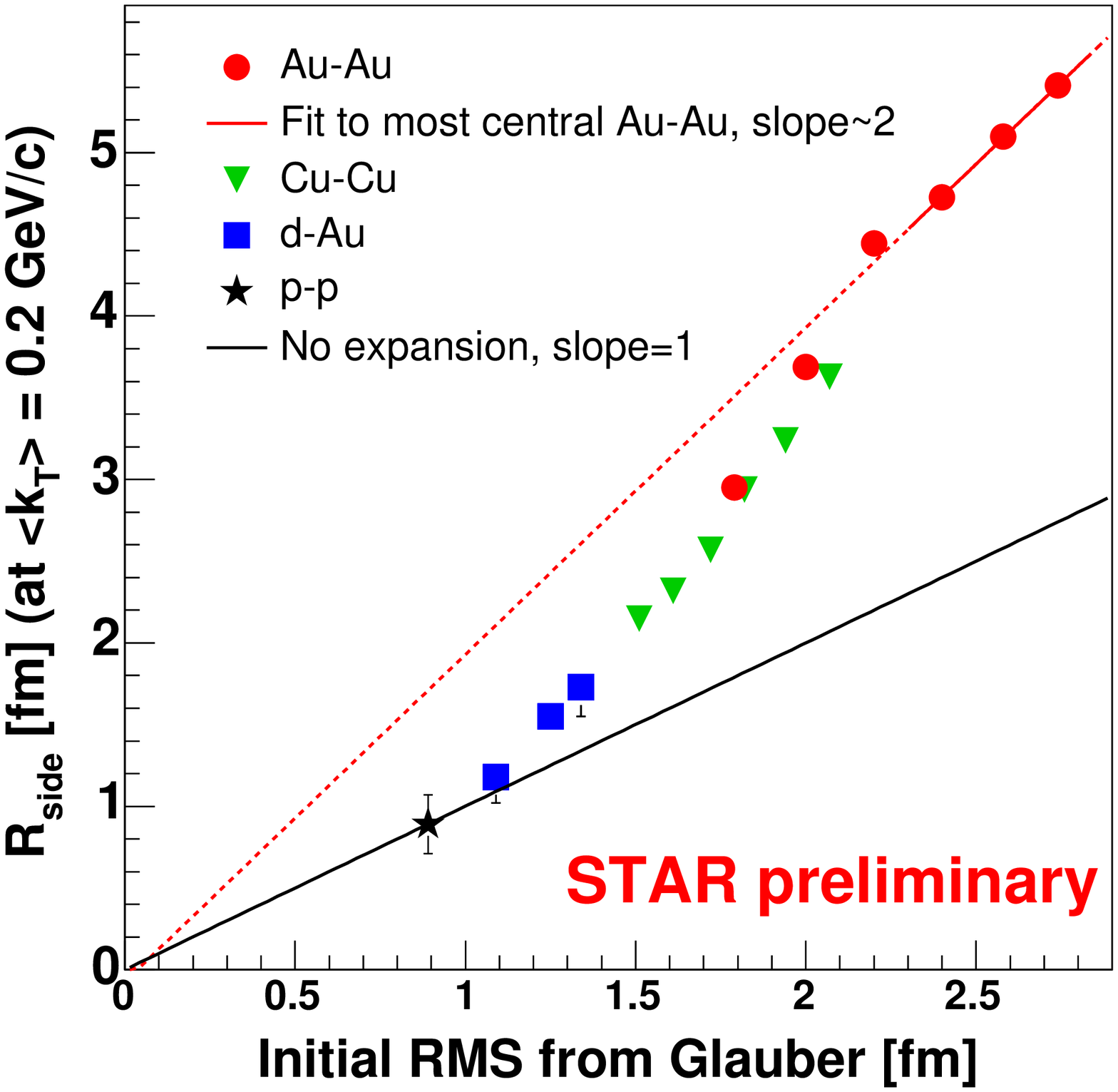}
   \vspace*{-2.3cm}
\end{minipage}
 \hspace{\fill}
 \begin{minipage}[t]{73mm}
   \vspace*{-0.3cm}
   \includegraphics[width=73mm]{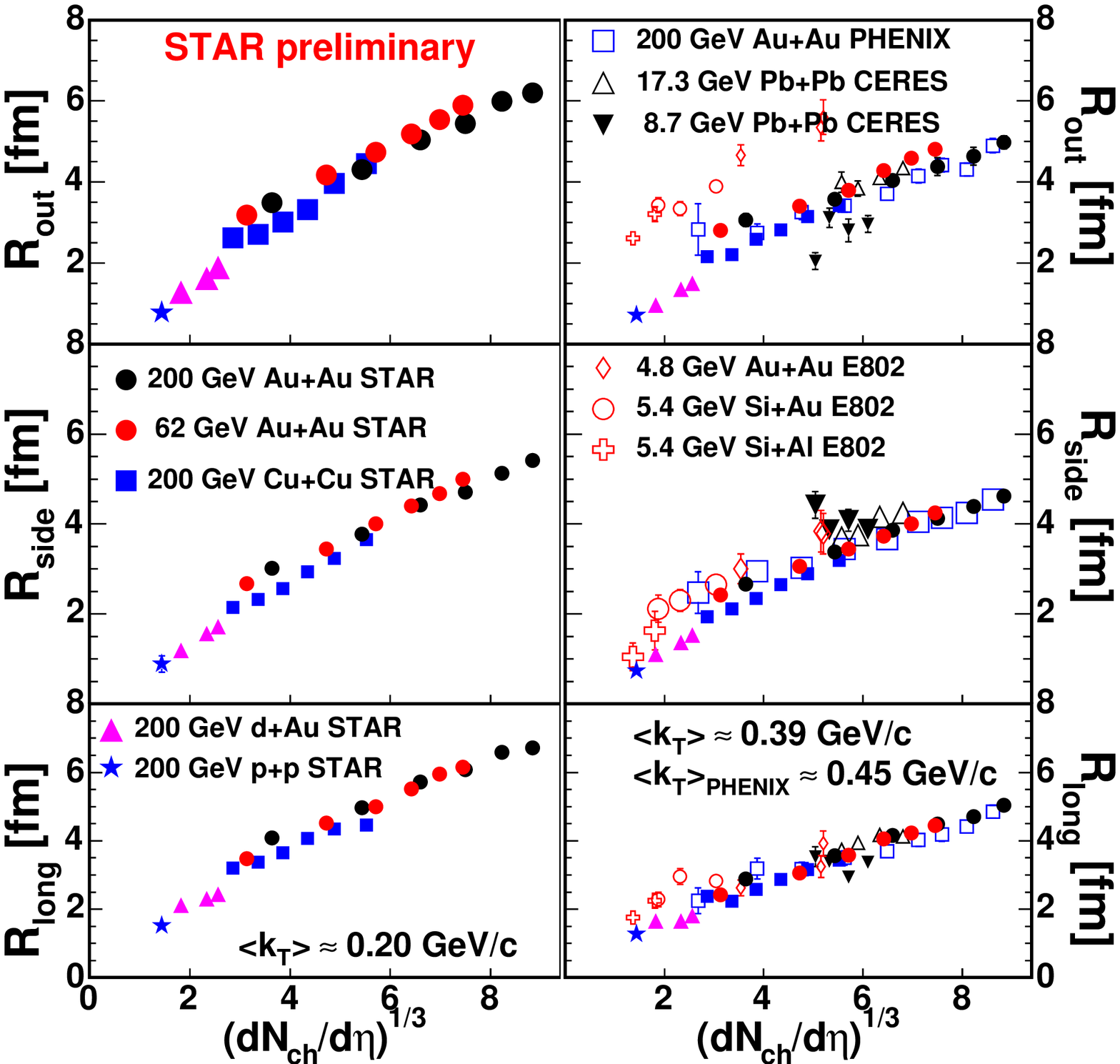}
   \vspace*{-0.5cm}
   \caption{a) Final size of the source vs. initial radii. STAR Au+Au data from \cite{Mercedes2004}; b) Femtoscopic radii dependence on the number of charged particles.}
   \label{fig:SystemExpansion}
 \end{minipage}
   \vspace*{-1.3cm}
\end{figure}

One of the differences that may be expected between such a large system as Au+Au 
and a small system as p+p is the expansion. This can be studied
in a model-dependent approach when the final RMS of the system is compared
to the initial one. The first value is equal, with good
accuracy, to $R_{side}$ calculated for the lowest $k_T$ range ($[0.15,0.25]$ GeV/c). The second one
is calculated with Glauber model for nuclei and a proton initial size was taken 
from an $e^-$ scattering \cite{eScattering} as a reference.
The result of this comparison for p+p, d+Au, Cu+Cu and Au+Au collisions at the 
same energy of the collision ($\sqrt{s_{NN}}$ = 200 GeV) are combined on the left panel of Figure \ref{fig:SystemExpansion}a.

As seen, the most central Au+Au collisions undergo an expansion by a factor of two
while p+p collisions show no or a little expansion. Data points from peripheral d+Au
collisions show similarity to p+p while central d+Au points exhibit an expansion 
like in peripheral Cu+Cu. Finally, central Cu+Cu expands similarly like peripheral
Au+Au. 

Figure \ref{fig:SystemExpansion}b presents the HBT radii 
dependence on $(dN_{ch}/d\eta)^{1/3}$ ($dN_{ch}$ - number of charged particles at midrapidity) for different colliding systems 
and at different energies of the collisions. The motivation for studying such  
a relation is its connection to the final state geometry 
through the particle density at freeze-out.  
All STAR results, from p+p, d+Au, Cu+Cu and  Au+Au collisions, are combined on the left panel
of this figure and, as seen, all radii exhibit a scaling with $(dN_{ch}/d\eta)^{1/3}$.
On the right panel of this figure STAR radii, this time for different 
$k_T$ range,
 are plotted together 
with AGS/SPS/RHIC systematics \cite{Lisa2005}. 
It is impressive that the geometric radii ($R_{side}$ and 
$R_{long}$) follow the same curve for different collisions  over a wide range 
of energies and, as it was checked, this observation 
is valid for all $k_T$ bins studied by STAR.
It is a  clear signature 
that the multiplicity is a scaling variable that drives geometric HBT radii at midrapidity. 
$R_{out}$ mixes space and time information. Therefore it is unclear
whether to expect the simple scaling with the final state geometry.
Although, because of the finite intercepts of the linear scaling \cite{Lisa2005,PhenixPRL2}, results
do not confirm predictions that freeze-out takes place at the constant density \cite{CsorgoCsernai}.
Additionally, this scaling was verified at midrapidity only, and some dependence on rapidity outside this region 
may be expected \cite{Stock,CsorgoCsernai} so it is not obvious if the scaling holds then.

As a result of this study, one can venture to predict the size 
of the source at midrapidity without knowing anything about the collision (like energy, $N_{part}$, 
 impact parameter, etc.) except for the multiplicity \cite{Stock,CsorgoCsernai,Lisa2005}. 
This scaling is expected to persist for all systems that are meson dominated
but is violated for  low energy collisions that are 
dominated by baryons \cite{Stock,Lisa2005,Ceres}.

\section{Transverse mass dependence of HBT radii}

\begin{figure}[t]
\begin{minipage}[t]{70mm}
   \vspace*{-0.3cm}
  \includegraphics[width=70mm]{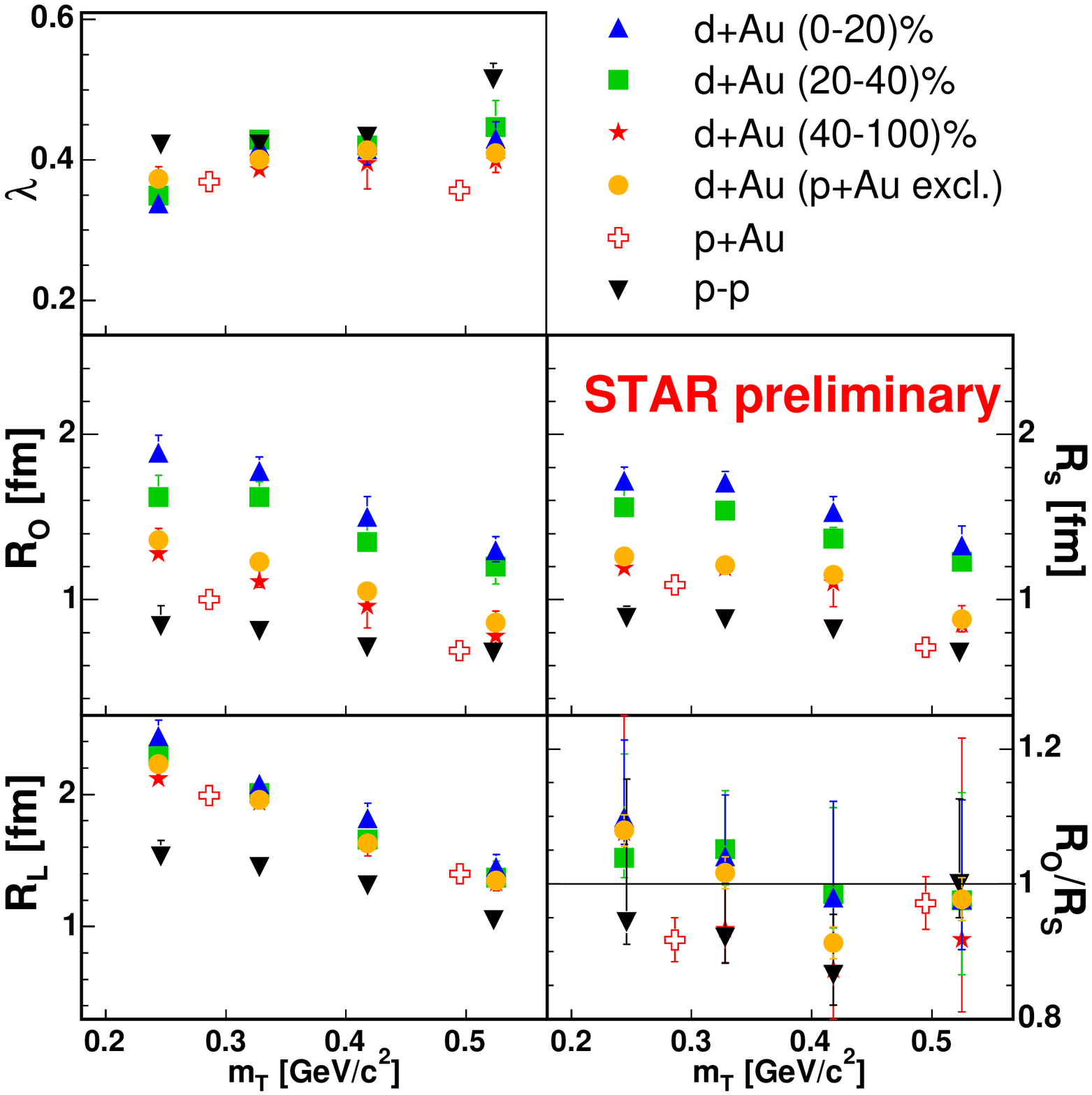}
\end{minipage}
 \hspace{\fill}
 %
 \begin{minipage}[t]{76mm}
   \vspace*{-0.3cm}
  \includegraphics[width=76mm]{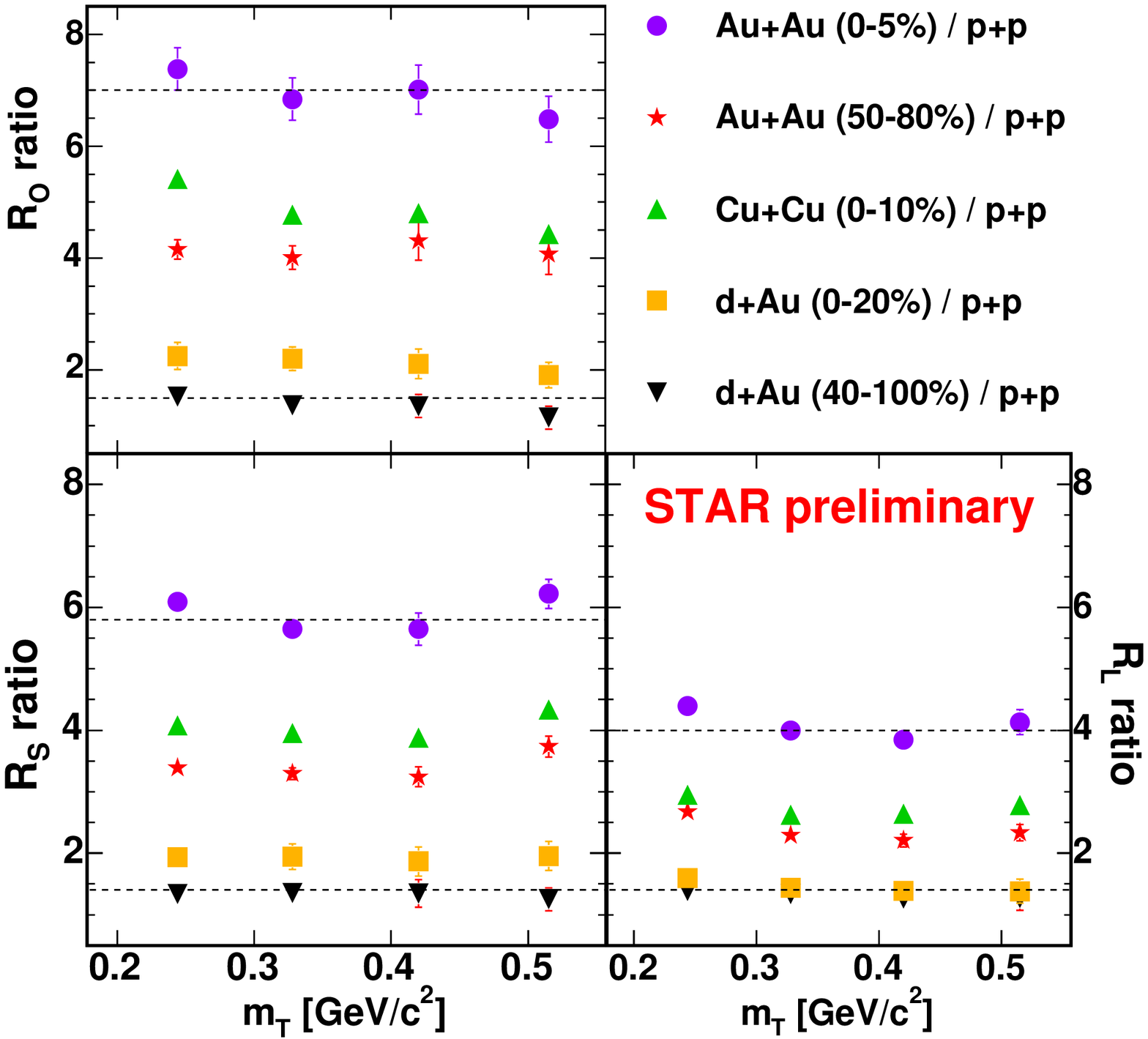}
\vspace*{-0.5cm}
   \caption{ a) $m_T$ dependence of HBT radii and $\lambda$ in p+p and  d(p)+Au  collisions at $\sqrt{s_{NN}}$ = 200 GeV; b) Ratio of HBT radii from Au+Au, Cu+Cu and d+Au by p+p collisions at $\sqrt{s_{NN}}=200$ GeV. }
   \label{fig:mTscaling}
 \end{minipage}
   \vspace*{-5.3cm}
\end{figure}

In heavy ion collisions a decrease of HBT radii with an increase of $m_T$ 
is commonly associated with the transverse flow of a bulk matter \cite{Mercedes2004}.  
Natural question would be whether this dependence looks different in
smaller systems like p+p or d+Au and what is the origin of this dependence.
On Figure \ref{fig:mTscaling}a the three dimensional radii from p+p and 
d(p)+Au collisions are plotted vs $m_T$.
For these systems femtoscopic sizes decrease with the increase of the transverse mass 
and d+Au results show also the dependence on the centrality like it is observed
in Au+Au collisions \cite{Mercedes2004}. 
Additionally, the value of $R_{side}$ and $R_{long}$ for p+Au collisions is similar 
to p+p collisions while $R_{long}$ is more like in d+Au collisions. 
Although it has to be emphasized that due to the way of extracting p+Au events from 
d+Au sample p+Au results correspond rather to peripheral p+Au collisions so the size of the
source is expected to be larger for central collisions. Therefore, results suggest
that the size of the source in p+Au collisions is not the same as in p+p.
Comparison of the peripheral d+Au collisions, that include about 15\% of p+Au collisions,
with and without extracted p+Au events
show no significant difference but that may be due to a fact that the d+Au sample still
includes n+Au events that cannot be excluded from the data. 


In elementary particle collisions, resonance production contributes significantly to the $m_{T}$ dependence
of the HBT radii \cite{Weiner}, while in heavy ion collisions, flow effects dominate this dependence \cite{Wiedemann1997}. 
The other scenarios 
that can give the similar dependence are the Heisenberg uncertainty
and the string fragmentation \cite{Alexander}.

On Figure \ref{fig:mTscaling}b the ratio of the three dimensional radii 
in Au+Au, Cu+Cu and d+Au collisions to p+p radii is plotted vs $m_T$.
Surprisingly, these ratios look flat although it is expected that
different origins drive the transverse mass dependences of the HBT radii
in Au+Au and p+p collisions. If these expectations are correct the data
show that one may not distinguish different physics between p+p and Au+Au collisions
studying pion interferometry.

An alternative explanation of this phenomena came from a work done by
Cs{\"o}rg{\H o} {\it et al.} \cite{Csorgo2004}.
Authors using a Buda-Lund hydrodynamic model, that successfully describes
the momentum correlations in Au+Au collisions, were able to 
fit STAR p+p spectra and HBT radii. But in this case they claim that
the transverse mass dependence of the femtoscopic sizes is not generated by
the transverse flow, but by the temperature inhomogeneities of hadron-hadron
collisions due to the freezing scale. Then the conclusion from this study 
would be that in p+p collisions the system has similar bulk properties as in Au+Au collisions.

 Non-identical particle correlations like $\pi$-K or
$\pi$-p in Au+Au collisions show a difference in then average emission points of two
particles that is due to flow \cite{STARkpi}. Therefore, femtoscopic study of particles with
different masses in p+p collisions could be used to verify a flow hypothesis in small
systems like p+p and d(p)+Au.

\section{Evidence of non-femtoscopic correlations}

When doing femtoscopic analysis in p+p and d+Au collisions a problem with
non-femtoscopic correlations has been observed. It is manifested
in a non-vanished tail of the correlation function to unity, for large $\vec{Q}$.
In elementary particle collisions \cite{Alexander, NA22} these non-femtoscopic correlations
were also observed and taken into account by adding an {\it ad-hoc} component to the parametrization of the 
correlation function that assumes that
the correlation function for large $\vec{Q}$ depends linearly on
the three components of the two-particle relative momentum (see equation \eqref{na22parametrization}).
\begin{equation}
C_2(Q_{out},Q_{side},Q_{long}) = (1+\lambda e^{-(Q_{out}^2R_{out}^2+Q_{side}^2R_{side}^2+Q_{long}^2R_{long}^2)})(1+\alpha Q_{out}+\eta Q_{side}+\zeta Q_{long}) \label{na22parametrization}
\end{equation}

\begin{figure}[t]
\begin{minipage}[t]{39mm}
\vspace*{-0.3cm}
  \includegraphics[width=39mm]{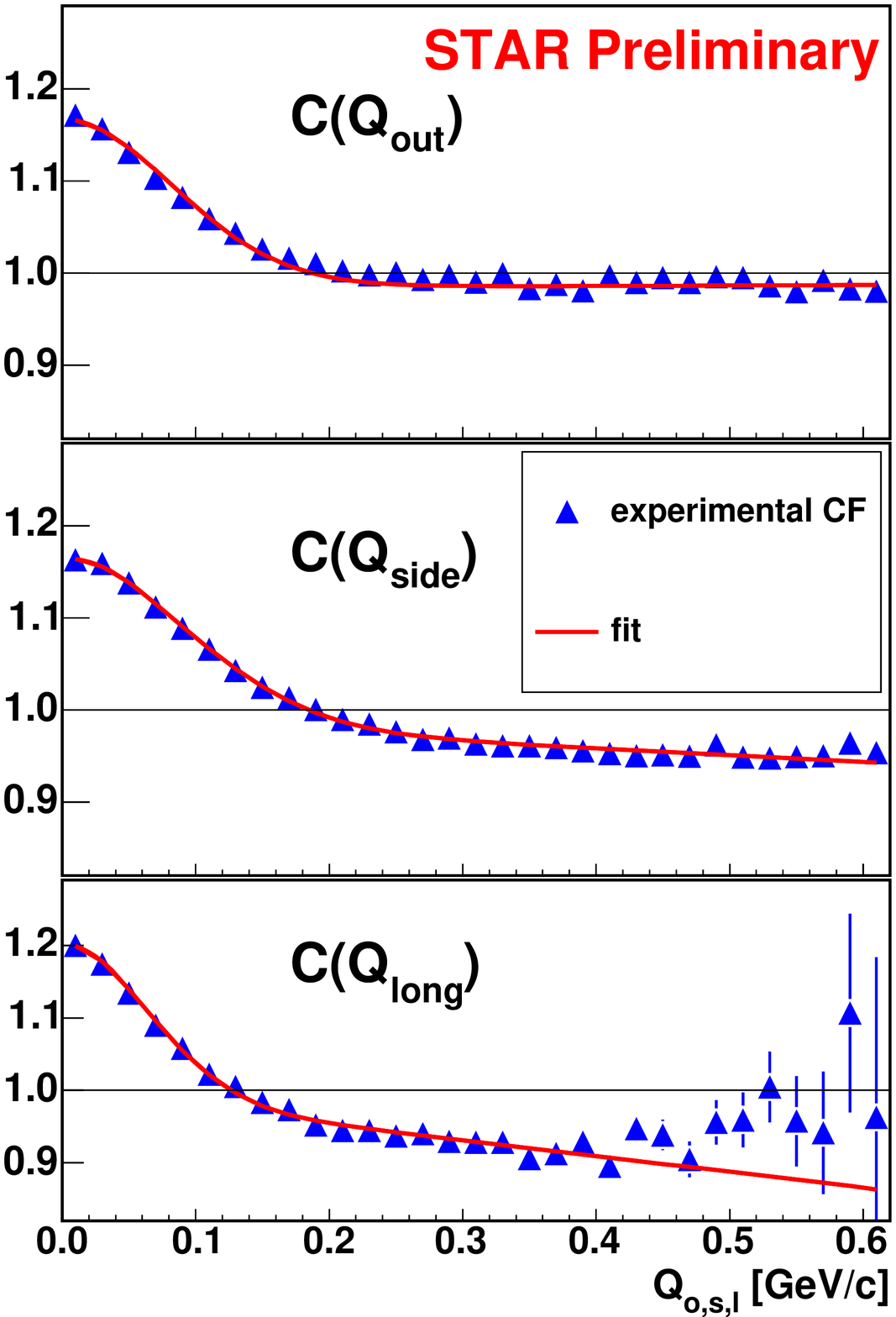}
\end{minipage}
\hspace{\fill}
\begin{minipage}[t]{107mm}
\vspace*{-0.3cm}
  \includegraphics[width=107mm]{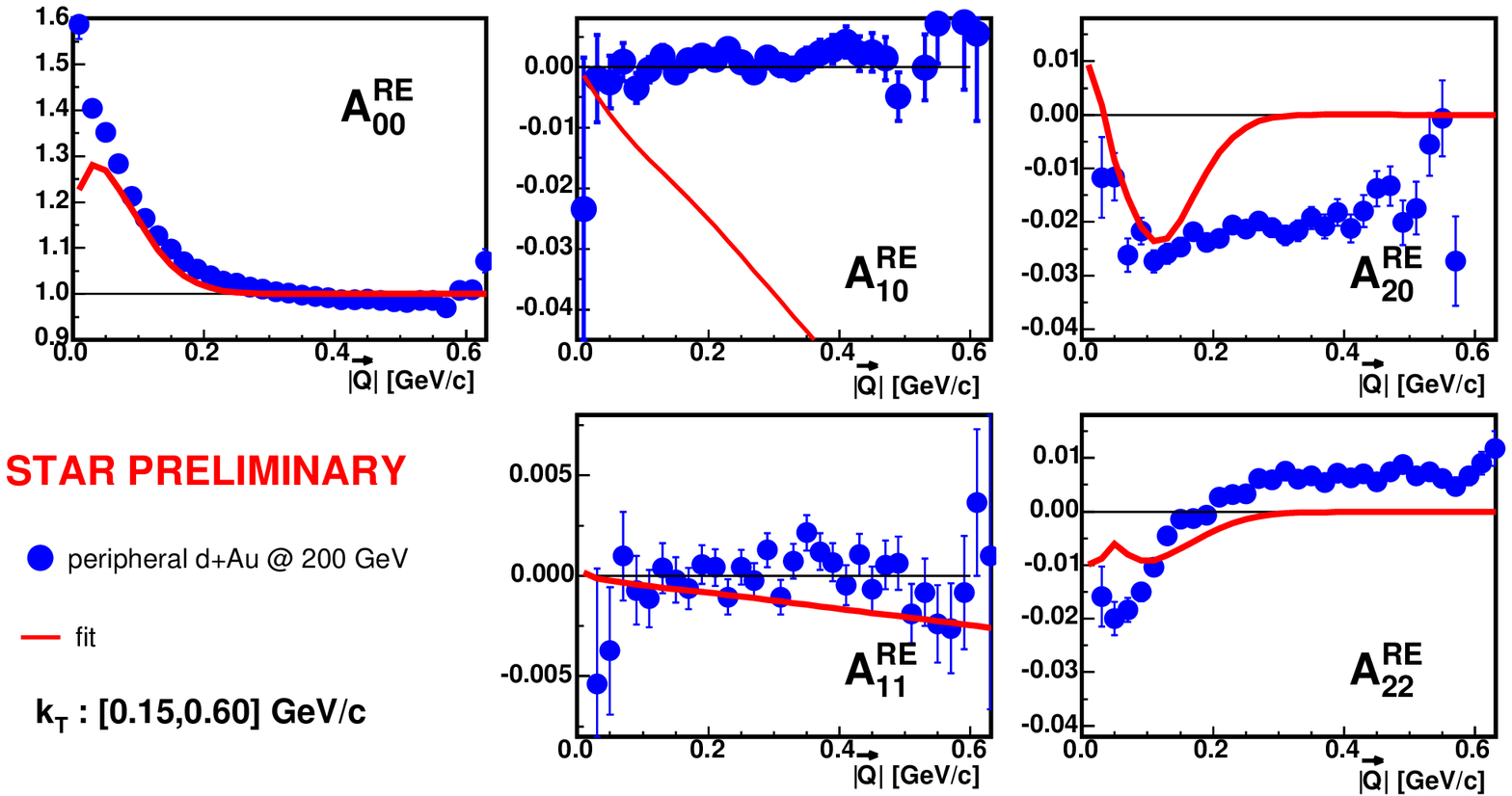}
\vspace*{-0.5cm}
  \caption{a) 1D projections of 3D correlation function. b) First five non-vanished components of the spherical harmonics decomposition of the correlation function. In a) and b) 3D correlations function for d+Au peripheral collisions was fitted with parametrization given by Eq.\eqref{na22parametrization}}
  \label{fig:na22}
\end{minipage}
\vspace*{-2.cm}
\end{figure}

Using this parametrization the fit to the STAR p+p and d+Au collisions was
performed and the femtoscopic radii turned out to be larger up to 30\% in comparison to 
the standard parametrization. Figure \ref{fig:na22}a shows
the projections of the 3D correlation function for the most peripheral d+Au collisions (that is STAR worst case)
and the projections of the fit described above. It looks like the fit matches
experimental data with good accuracy 
but more careful study is required to judge on the correctness of the new parametrization and will
be performed with method described below.

A common approach to present 3D correlation function is to project it onto
the three components of $\vec{Q}$ separately, as shown on Figure \ref{fig:na22}a.
The disadvantage of this approach
is that when doing such projections one has to constrain non-projected components 
to keep a signal but then the full information on the correlation function in the 3D space is lost.
To eliminate this inconvenience a new approach of studying correlations was applied
that is based on a decomposition of the correlation function into spherical harmonics (for detailed description of this method see \cite{Chajecki2005}).
In this method no cuts are performed on $\vec{Q}$'s components
what allows to recognize symmetries in $\vec{Q}$-space to see, looking at 1D plots,
relevant aspects of 3D correlation functions.

Figure\ref{fig:na22}b shows the first few 
components of the decomposition of the correlation function onto the spherical harmonics 
for peripheral d+Au collisions. The fitted correlation function, that includes a new term  to account 
for non-femtoscopic effect, was decomposed using this method.
As shown on Figure \ref{fig:na22}a the new parametrization fits the correlation function with good accuracy 
but with the spherical harmonic method it is seen that the fit is not correct.
Distributions for $l$=1 are non-zero and $A_{1,0}$ shows a strong dependence on $|\vec{Q}|$.
In a  system like $\pi-\pi$ at midrapidity all odd components should vanish by symmetry. 
Additionally, the new parametrization does not fit the baseline of the correlation function 
that has an evidence in non-vanished $A_{2,0}$ and $A_{2,2}$ distributions for large $|\vec{Q}|$.

This study shows that it is not sufficient to look at the Cartesian projections of the correlation function
to judge about the quality of the fit and the correctness of
the used parametrization. It is required to use spherical harmonic method to see
the experimental data and the fit in the 3D space.

The analytic formulas of spherical harmonics are well-known so the $A_{2,0}$ and $A_{2,2}$ distributions 
may be parametrized and included in the fit.
Such study was presented in \cite{ChajeckiQM05} and it showed a good agreement
with experimental data.  Due to the lack of the space the results are not presented here,
but the radii in p+p and d+Au collisions 
are changed up to 10\% the most, although the $m_T$ scaling described in the previous section persists.

\section{Conclusions}

The results of pion interferometry for several energies and colliding systems at RHIC
 have been presented. 
In agreement with data at SPS and AGS, STAR 
indicates that the multiplicity is the scaling variable that determines 
the size of the source at freeze-out at midrapidity.
The $m_T$ dependence of HBT radii seems to be independent of 
collision species or multiplicity.
Finally, a problem with the baseline of the correlation function for low multiplicity 
collisions has been reported, and a promising tool based on the spherical 
harmonic decomposition of the correlation function has been used in order to address it.
The physics of this structure remains under investigation.
The advantage of this method in studying the correctness of the parametrization of the
correlation function used in a fit has been shown.



\begin{thebibliography}{9}

\bibitem{Lisa2005} M. Lisa, S. Pratt, R. Solz, U. Wiedemann, Annu. Rev. Nucl. Part. Sci. (2005) 55:357-402. 

\bibitem{TPC} M. Anderson {\it et al.}, Nucl. Instrum. Meth. A 499 (2003) 659 .

\bibitem{Pratt90} S. Pratt, T. Cs\"{o}rg\"{o}, and J. Zimanyi, Phys. Rev. C 42 (1990) 2646.
\bibitem{Bertsch89} G. Bertsch, Nucl. Phys. A 498 (1989) 173c.


\bibitem{Bowler} M. G. Bowler, Phys. Lett. B 270 (1991) 69.
\bibitem{Sinyukov} Yu. M. Sinyukov {\it et al.}, Phys. Lett. B 432 (1998) 249.

\bibitem{PhenixPRL2} S.S. Adler {\it et al.} (PHENIX Collaboration), Phys. Rev. Lett. 93 (2004) 152302.
\bibitem{CsorgoCsernai} T. Cs{\"o}rg{\H o} and L. P. Csernai, Phys. Lett. B 333 (1994) 494.
\bibitem{Stock} R. Stock, Annalen der Physik, 48 (1991) 195.
\bibitem{Ceres} D. Adamo\'va {\it et al.} (CERES Collaboration), Phys. Rev. Lett. 90 (2003) 022301.

\bibitem{Mercedes2004} J. Adams {\it et al.} (STAR Collaboration), Phys. Rev. C 71 (2005) 044906.
\bibitem{eScattering} I. Sick, Eur. Phys. J. A 24, s1, (2005) 65-67.

\bibitem{Weiner} R.M. Weiner, Phys. Rep. 237 (2000) 249-346.
\bibitem{Wiedemann1997} U. A. Wiedemann, U. Heinz, Phys. Rev. C 56 (1997) 3265.
\bibitem{Alexander} G. Alexander, ArXiv:hep-ph/0302130.
\bibitem{Csorgo2004} T. Cs{\"o}rg{\H o}, M. Csan{\'a}d, B. L{\"o}rstad, A. Ster, ArXiv:hep-ph/0406042.
\bibitem{STARkpi} J. Adams {\it et al.} (STAR Collaboration), Phys. Rev. Lett. 91 (2003) 262302.
\bibitem{NA22} N.M. Agababyan {\it et al.} (NA22 Collaboration), Z. Phys. C71 (1996) 405.
\bibitem{Chajecki2005} Z. Chaj\c{e}cki {\it et al.} (for the STAR Collaboration), ArXiv:nucl-ex/0505009.
\bibitem{ChajeckiQM05} Z. Chaj\c{e}cki (for the STAR Collaboration), {\it Proceedings of the Quark Matter 2005 Conference}, to appear in Nucl. Phys. A, ArXiv:nucl-ex/0510014.


\end{thebibliography}
\end{document}